\documentstyle[12pt,aaspp4,flushrt,psfig]{article}
\begin{document}

%

\title{MEASUREMENT OF THE GALACTIC X-RAY/$\gamma$-RAY
BACKGROUND RADIATION: CONTRIBUTION OF DISCRETE SOURCES} 

\author{
Azita Valinia\altaffilmark{1,2}, Robert L. Kinzer\altaffilmark{3}, and  
Francis E. Marshall\altaffilmark{1}}  

\altaffiltext{1}{Laboratory for High Energy
Astrophysics, Code 662, NASA/Goddard
Space Flight Center, Greenbelt, MD 20771; valinia@milkyway.gsfc.nasa.gov}
\altaffiltext{2}{Department of Astronomy, University of Maryland,
College Park, MD 20742}
\altaffiltext{3}{E. O. Hulburt Center for Space Research, Naval Research
Laboratory, Washington, DC 20375-5352}

\begin{center}
Accepted for Publication in the Astrophysical Journal
\end{center}

\begin{abstract}
The Galactic background radiation near the Scutum Arm 
was observed simultaneously with RXTE and OSSE in order to determine 
the spectral shape and the origin of the emission in the  
hard X-ray/soft $\gamma$-ray band. The spectrum in the 3 keV to 1 MeV band 
is well modeled by 4 components: a high energy continuum dominating
above 500~keV that can be characterized by a power law of photon index
$\sim 1.6$ (an extrapolation from measurements above $\sim 1$~MeV);
a positron annihilation line at 511~keV and 
positronium continuum;
a variable hard X-ray/soft $\gamma$-ray component that dominates 
between 10-200~keV
(with a minimum detected flux of 
$\sim 7.7 \times 10^{-7} \,{\rm photons \, cm^{-2} \, s^{-1} \, 
keV^{-1} \, deg^{-2}}$ at 100~keV averaged over the field of view of
OSSE) and that is well modeled by   
an exponentially cut off power law of photon index $\sim 0.6$
and energy cut off at $\sim 41$~keV; 
and finally a thermal plasma model of
solar abundances and temperature $\sim 2.6$~keV that dominates below
10~keV. We estimate that the contribution of bright
discrete sources to the minimum flux detected by OSSE was $\sim 46\%$
at 60~keV and $\sim 20\%$ at 100~keV. 
The remaining unresolved emission may be interpreted 
either as truly
diffuse emission with a hard spectrum (such as that from inverse
Compton scattering) or the superposition of 
discrete sources that have
very hard spectra. 
\end{abstract} 
 

\newpage
%
 
\section{INTRODUCTION}
Since its discovery, the Galactic X-ray/$\gamma$-ray background,
particularly from the ridge (i.e. the narrow region centered on the plane 
covering approximately $\pm 60^0$ in longitude), 
has been studied with every major X-ray and $\gamma$-ray observatory. 
The spectrum of the emission is reasonably well measured and understood 
above $\sim 1$~MeV (e.g., Kinzer, Purcell, \& Kurfess 1999; 
Strong et al. 1996a; Bloemen et al. 1997; Hunter et al. 1997; Hunter, Kinzer, \&
Strong 1997). At energies
above 100~MeV, the dominant emission process is the decay 
of $\pi^0$ meson produced
in the interaction of cosmic ray
nucleons with the interstellar matter (e.g., Bertsch et al. 1993).
Between $\sim 1$ and 70~MeV, electron bremsstrahlung and inverse Compton
scattering appear to dominate over discrete sources
(e.g., Sacher \& Sch$\ddot{\rm o}$nfelder 1984;
Kniffen \& Fitchel 1981, Skibo 1993). 
However, in the hard X-ray/soft $\gamma$-ray band (3-500~keV) 
the shape of the spectrum and the origin of the emission remain
uncertain. 
At soft $\gamma$-ray energies (below 1~MeV), multiple components
are believed to contribute to the total emission. These
include transient discrete sources, positron annihilation
line and 3-photon positronium continuum radiation, and 
a soft $\gamma$-ray component dominant up to about 300~keV of
unknown origin. This component, measured with the CGRO's Oriented Scintillation
Spectrometer Experiment (OSSE), can be roughly
characterized by simple power law models of indices between 2.3 and 3.1
at different locations on the Galactic plane
(e.g., Kinzer et al. 1999; Skibo et al. 1997).
More recently, the soft $\gamma$-ray emission from the Galactic center
was measured by the HIREGS balloon-borne germanium spectrometer 
and was characterized by 
a single power law of photon index $\sim 1.8$ 
plus the positronium component (Boggs et al. 1999).
However, many of the soft $\gamma$-ray observations, particularly those from the
central region of the Galaxy, are contaminated by bright
and variable discrete sources.
At hard X-ray energies ($10-35$~keV), the overall spectrum of the Galactic plane
background was characterized by a power law 
of photon index $\sim 1.8$ with RXTE (Valinia \& Marshall 1998;
hereafter VM98). 
Hard X-ray emission above 10~keV was also detected with Ginga (Yamasaki et al.
1997). 

How the spectral shape of the background radiation 
extends from the hard X-ray to the soft $\gamma$-ray regime and 
how much of the emission is due to discrete sources remains to be
determined and is the subject of this paper.
Determining the exact nature of the spectrum in this band has significant implications for
the energetics of the Interstellar Medium (ISM). 
For example, a power law
spectrum extending from 10~keV to 1~MeV, if interpreted to be of diffuse 
nonthermal origin, has been proposed to result from nonthermal electron 
bremsstrahlung (e.g. Skibo et al. 1997).  
However this process is energetically very demanding since electron
bremsstrahlung at these energies is highly inefficient.
A power of $10^{42}-10^{43}\,{\rm erg \,s^{-1}}$ is required, which approaches
or even exceeds the power injected into the Galaxy via supernovae explosions
(Skibo et al. 1997).
Attempting to explain the nature of the emission in terms of diffuse thermal
processes is equally unsatisfactory because plasma temperatures of
80-100 keV are implied. Since the gravitational potential of the Galaxy is
only on the order of 0.5 keV, it is not clear how such a plasma would
be generated and confined to the Galactic plane.

Unfortunately, measurement of the Galactic background radiation in 
the hard X-ray/soft
$\gamma$-ray band and its interpretation is inherently
difficult with current instruments because of the presence of   
numerous transient, hard discrete sources in the 
Galactic plane, and the fact that generally hard X-ray/$\gamma$-ray 
instruments either have large
fields of view and no imaging capabilities 
or have imaging capability but no
diffuse emission sensitivity. As a result,
distinction of diffuse emission
from point sources with current instruments has remained a difficult task.
For this reason, simultaneous, multiple instrument observations
are necessary.
To date, coordinated observations of the Galactic center region with 
OSSE and the imaging instrument SIGMA has been performed (Purcell et al. 1996).
However, SIGMA has a sensitivity of about 25~mcrab ($2\sigma$) for a 
typical 24 hour observation. As a result, weak sources 
escape detection in such a survey and the unresolved spectrum can still
be significantly contaminated by discrete source contribution. 

In this paper, we present 
contemporaneous observations of the Galactic
background emission near the Scutum Arm (centered at $l=33^\circ$) 
over the 3~keV to 1~MeV range with RXTE and OSSE. RXTE has a relatively
small field of view of $1^\circ$ FWHM with hard X-ray capability and
discrete source sensitivity of $\sim 1$~mcrab in the 2-10~keV band, making
the detection of hard, discrete sources in the field of view of OSSE possible. 
OSSE, with a 
field of view of $11^\circ.4 \times 3^\circ.8$ FWHM, is sensitive to
diffuse emission in the soft $\gamma$-ray band.  
The Scutum Arm was 
chosen because it exhibits bright and apparently diffuse emission 
(e.g. Kaneda 1997; VM98), and  
unlike the Galactic center region for example, there are few bright discrete
sources in the field of view.
This direction is approximately tangent to the 5-kpc arm of
our Galaxy as seen from the Earth. The arm contains large numbers of young
stars (Hayakawa et al. 1981), and an unusual concentration of high-mass X-ray
binary transients (van Paradijs 1995).
Our main goal is to constrain the shape of the spectrum 
and understand its origin in the $10-400$~keV band. 
In \S 2 we present our observations and in \S 3 we present the results of
our analysis. 
Finally in \S 4 we discuss their implications.

\section{OBSERVATIONS} 

The simultaneous RXTE and OSSE observations were performed from January 28
through February 23
of 1998. The OSSE instrument (Johnson et al. 1993) consists
of four Na(TI)/CsI(Na) phoswich detectors which cover the energy range 
50 keV-10 MeV. 
It has an effective area of $\sim 2000 \, {\rm cm}^2$ 
and average energy resolution of $\sim 8.8\%$ at 511~keV.  
The field of view is approximately rectangular with FWHM of
$11^\circ.4 \times 3^\circ.8$.
During the 4 weeks of observation, OSSE (with its long collimator 
axis parallel to the Galactic plane) was continuously oriented toward 
Galactic coordinates $(l,b)=(33^\circ,0^\circ)$ 
except for alternating 2-minute intervals during which offset background 
measurements were made. 
The background was taken alternately at 
Galactic latitudes $\pm9^\circ$ ($l=33^\circ$) during weeks 1 and 3 
of the observation and
$\pm12^\circ$ ($l=33^\circ$) during weeks 2 and 4 of the observation.

The RXTE observations were planned such that during
the 4 week observation, RXTE was 
scanning the full field of view of OSSE 
(i.e. the region $44^\circ < l < 22^\circ$
and $-4^\circ < b< 4^\circ$) every day for either one or two hours
(exposures alternated between one and two hour intervals). 
The goal of these observations were two-fold. One was 
to monitor discrete sources in the field 
of view of OSSE
and to estimate their contribution to OSSE's flux.
The second goal was to measure the spectrum of the diffuse emission in 
the hard X-ray band and simultaneously model that with the spectrum
measured with OSSE. 
The RXTE scans were performed with the Proportional Counter 
Array (PCA). The PCA (Jahoda et al. 1996) has a total collecting area
of 6500 ${\rm cm}^2$, an energy range of 2-60 keV, and energy
resolution of $\sim 18\%$ at 6 keV. 
The field of view of the collimator is 
approximately circular with FWHM of $1^\circ$.  
The RXTE ``diffuse'' emission spectrum was obtained by
scanning the field of view of OSSE
excluding the regions that the scans went
over known and detected discrete sources. 
The most recent PCA background estimator program 
{\it pcabackest} (version~2.0c; L7 model) provided by the
RXTE Guest Observer Facility was used 
to estimate the PCA background.  

\section{ANALYSIS}

Figure~1 shows the composite diffuse plus discrete-source 
emission spectrum as measured by OSSE in
weeks 1 and 2 (filled circles - hereafter W1-2) and weeks 3 and 4 
(open circles- hereafter W3-4) of the observations. 
The lowest energy data point for each
observation has an approximately 20\% systematic uncertainty
and was therefore not used for modeling the data. 
In order to convert the flux through the OSSE collimator to 
the diffuse flux per radian of the Galaxy, 
a $5^\circ$ FWHM Gaussian distribution in latitude and constant intensity
in longitude for the spatial distribution of the emission was assumed
(e.g., Kinzer et al. 1999; Purcell et al. 1996).  
In the 8-35~keV band, the FWHM of the distribution was derived to be
$4^\circ\!.8^{+2.4}_{-1.0}$ with RXTE (VM98). 
The converted flux per radian is therefore a function of 
this latitude assumption. For example, assuming a
FWHM of $2^\circ$ would lower the flux per radian by 
$\sim 30\%$.
Assuming a FWHM of $8^\circ$ would increase the flux per radian by
$\sim 50\%$ (see Figure~4 of Kinzer et al. 1999).
We will discuss how this assumption affect the spectral fits in \S 3.2. 

\begin{figure}[h]
\centerline{ 
{\hfil 
\psfig{figure=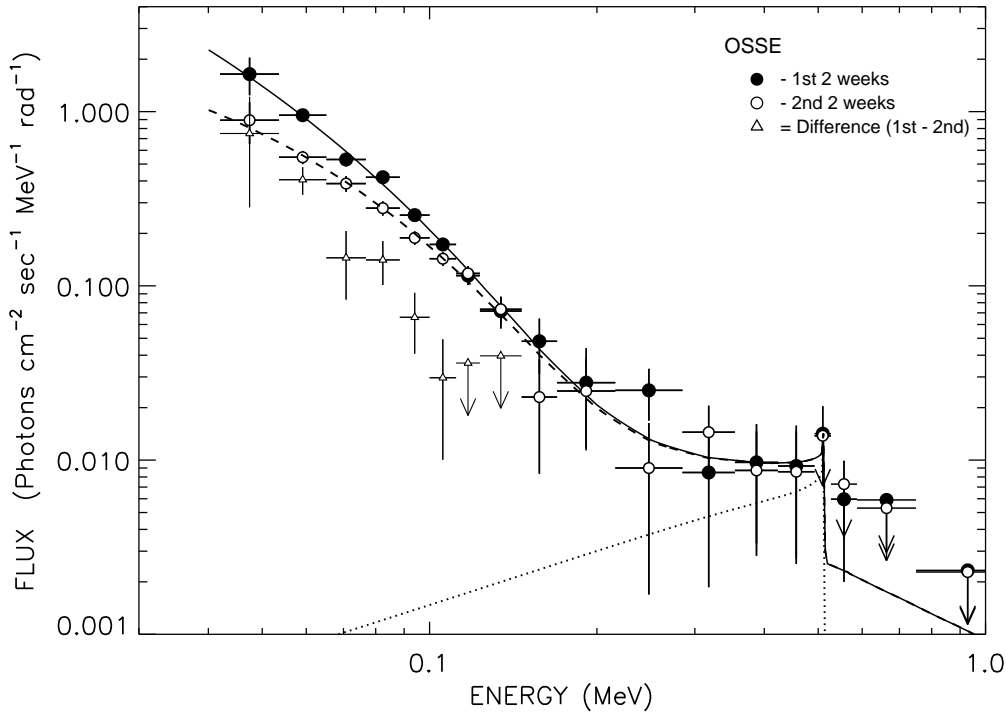,height=4.0truein,angle=90.0}
\hfil}
}

\caption
{\footnotesize Emission spectrum as measured by OSSE in weeks 1 and 2 (filled circles)
and weeks 3 and 4 (open circles). Diffuse emission of Gaussian distribution
with FWHM of $5^{\circ}$ in latitude and constant in longitude has been
assumed for OSSE's response. The spectrum can be modeled by a composite
of 3 components: a soft $\gamma$-ray component; positron annihilation
line (at 511~keV)
and 3-photon positronium continuum; and extrapolation of the
high energy continuum from above 0.5~MeV.
The solid and dashed line indicate the overall composite best fit models in
the first half and second half of observations, respectively. The
dotted line indicates the positron annihilation component
from the full 4-week observation. Note that the lowest energy point for
each observation has an approximately 20\% systematic uncertainty and
was not used in modeling the data.
}
\end{figure}

As seen from Figure~1, the spectra in the 40 keV to 1 MeV range  
from the two viewing periods are 
similar except in the $\sim 40-100$~keV energy range where variable discrete
sources have apparently altered the shape of the spectrum. 
The intensity in the $60-100$~keV band
dropped by about 32\% between the two viewing periods. 
The difference spectrum 
of the two viewing periods (triangles in Figure~1) can be 
modeled by a power law of photon 
index $\sim 4.3$. 
In what follows, we first discuss the detection and estimated contribution
of discrete sources to the total OSSE flux. We then discuss the 
modeling and characteristics of the measured RXTE/OSSE spectra. 

\subsection{Discrete Sources}

Table~1 lists bright X-ray sources detected during the PCA scans  
that were within the field of view of OSSE. 
It also includes GRS~1915+105,
which is near the edge of the field of view,
since it is known to be extremely variable with intensities
as high as a few Crabs (e.g., Muno et al. 1999). 
The first four sources are accretion-driven pulsars while the last two
sources are microquasars. 
The spectra of these sources could
not be well determined above 20~keV with our RXTE observations because the 
scans had insufficient exposure time. 
For each of these sources, we determined a spectral
shape from other reported observations and estimated their 60 and 100~keV
flux by normalizing
their spectra with the average 2-10~keV flux during the PCA scans.  
We used public RXTE data for XTE~J1858+034 (from observations
on Feb. 20 and 24, 1998) 
and GS~1843+009 (from March 5, 1997).
For both of these sources we model the photon distribution
using the standard form for pulsars 
\begin{equation}
A(E)=k(E/1\,{\rm keV})^{-\Gamma}\, \exp((E_c-E)/E_f) 
\end{equation}
(e.g. White, Swank, \& Holt 1983). For XTE~J1858+034, we find
$\Gamma \sim 1.3$, $E_c \sim 2.3$ and $E_f \sim 24.6$. For GS~1843+009,
we find $\Gamma \sim 0.6$, $E_c \sim 13.3$ and $E_f \sim 22.3$.
Koyama et al. (1990) found similar parameters
($\Gamma \sim 0.71$, $E_c \sim 18.3$ and $E_f \sim 25$)
during the 1988 observations with Ginga when the source intensity
varied between 30 and 60 mCrabs.
In the case of A1845-024 (also identified as GRO~1849-03), we used
the BATSE spectrum of this source during outburst 
reported by Zhang et al. (1996) characterized by a power law of 
$\Gamma \sim 2.8$. 
For XTE~J1855-026, we used the RXTE spectrum reported by Corbet et al. (1998)
characterized by the pulsar model (equation~1) and parameters
$\Gamma \sim 1.23$, $E_c \sim 14.7$ and $E_f \sim 27$.  
For SS433, we used an exponentially cutoff power law model of 
photon index $1.5$ and cutoff energy at $20$~keV (Band, 
private communication, 1999). During our observations, GRS1915+105 was 
in a a very soft spectral state with an average power law photon 
index of $\sim 5.7$ and was not detected with RXTE above 40~keV (Muno \& Morgan
1999, private communication; Heindl 1999, private communication).

According to our estimate, the integrated 60-100~keV flux of the combined 
known bright sources listed in Table~1 (as would be 
seen through the OSSE's collimator)
decreased by about 41\% (from 
$(1.22\pm 0.12) \times 10^{-3}$ to $(7.14\pm 0.66) \times 10^{-4}$~${\rm 
photons \, cm^{-2} \, s^{-1}}$) from the first viewing period (W1-2) to
the second viewing period (W3-4) while the total OSSE measured flux for the entire
field of view dropped by 32\% (from $(3.68\pm 0.15) \times 10^{-3}$ to 
$(2.50\pm 0.11) \times 10^{-3}$~${\rm photons \, cm^{-2} \, s^{-1}}$).   
It appears that the unresolved portion of the emission 
(i.e. total OSSE flux minus the estimated 
contribution of discrete sources flux) has 
also decreased from the first to the second viewing period. 
We offer two plausible explanation for this. One is that 
if indeed, the residual
flux is made of discrete sources, this implies that the
intensity of the integrated unresolved sources as seen through
the OSSE collimator has also decreased. 
There may be a tendency to believe that the integrated contribution 
of hundreds of discrete sources cannot
change very much. However, the integrated contribution can easily be
dominated by one or two sources. If these sources are highly variable as
shown in Table 1, then the integrated flux can also vary substantially.
Furthermore, RXTE's total monitoring time amounts to only $\sim 7$\% of
the total observation time by OSSE. If these sources showed substantial
variability on the order of a day, it is possible that a potential
source outburst escaped detection by RXTE but its flux was
continuously measured with OSSE. 
Another explanation is that the apparent decrease in the unresolved flux
is due to inaccuracies in our estimates of the contribution of known sources
to the emission. These estimates are unavoidably uncertain because they
use spectral fits from other observations and the spectral fits are
generally at energies below 50 keV. 
Consequently spectral variations with time or deviations from the
assumed spectral model will lead to errors in the estimated fluxes.
In particular, since the 2-10 keV fluxes are generally lower during
the second viewing interval, luminosity dependent spectral changes
will cause systematic differences in the estimated fluxes for the
two observing intervals.
For example, from Table~1 it appears
that GS~1843+009 is the dominant contributor among discrete sources.
During the first viewing period, the 2-10~keV flux of this source was a
factor of 2.7 lower than that measured during its bright state observed on
March 5, 1997, and the flux was a factor of $\sim 5$ lower during the
second viewing period. We have assumed the same spectral shape for both
observations. We are not aware of a comprehensive study documenting the
relation between luminosity and spectral shape of pulsars, but there is some
evidence that the spectrum of accretion-driven pulsars depends on
luminosity. Reynolds, Parmar, \& White (1993) found that the spectrum at
low energies became harder and $E_c$ decreased as the luminosity of the
transient pulsar EXO~2030+375 decreased during an outburst. As a result,
the ratio of the extrapolated hard X-ray (50-100 keV) flux to the 2-10 keV
flux decreased as the source luminosity decreased. The hardness ratio
``HR'', defined as the ratio of the flux at 50 keV to the flux at 5 keV,
decreased by $\sim 2$ as the luminosity decreased by $\sim 25$ from $1.0
\times 10^{38} \,{\rm ergs \, s^{-1}}$. 
On the other hand, Koyama et al.
(1990) did not find luminosity dependent spectral changes for GS~1843+009
while the source varied by a factor of $\sim 2$. 
If the spectrum of
GS~1843+009 during the OSSE measurements is softer than the spectrum
measured when it was more luminous, then its contributions to the OSSE
measurements have been overestimated, particularly in the second viewing
period when the source luminosity was a factor of 5 lower than the bright
state observation of March 5, 1997. A factor of 2 decrease in the
contribution of GS~1843+009 would increase the residual flux at 60 keV
from $5.8 \times 10^{-5}$ to $7.2 \times 10^{-5} \, {\rm photons \,
cm^{-2} \, s^{-1}}$ for the second observing interval. This decreases the
difference in the residual fluxes for the two viewing intervals (the
residual flux at 60~keV for the first viewing period was determined to be
$10.1 \times 10^{-5} \,{\rm photons \, cm^{-2} \, s^{-1}}$ from Table~1).
The fact that Koyama et al.
(1990) did not find luminosity dependent spectral changes for GS~1843+009
makes this explanation less compelling.
We also note that the HR for
our spectral model for GS~1843+009 is $\sim 3$ times larger
than that of the Koyama et al. model and is also larger than
that of any of the pulsars in the review of White, Swank, and Holt (1983).
This suggests that the values in Table~1 may, in fact, be overestimates of
the contributions of GS~1843+009.

Because of these uncertainties, we do not subtract the
contribution of these sources from the OSSE spectrum reported in the
next section but 
instead, we model {\it only} the 
OSSE data during the {\it second} viewing period when the intensity 
of discrete sources was the least. 

\subsection{Spectral Characteristics}

\subsubsection{OSSE data: 50-400 keV}

We now focus on the spectral characteristics of the soft $\gamma$-ray 
Galactic background emission.
As discussed by Kinzer et al. (1999), 
the gamma ray continuum between 
50 keV and 10 MeV can be described by a composite
of 3 independent components: 
(1) a soft $\gamma$-ray component dominant up to 
about 300 keV of uncertain origin; 
(2) a hard $\gamma$-ray component (hereafter high energy
continuum or HE) which is the extrapolation  of the HE component
above 1~MeV and is likely due to the interaction of cosmic rays 
with the ISM dominating above 500~keV (e.g. Skibo 1993); (3)   
positron annihilation line and 3-photon positronium 
continuum radiation (hereafter PA; Ore \& Powell 1949) 
which are strongly enhanced toward the Galactic center. 
Fits to the OSSE data accumulated during the entire 4 week observation
were used to determine the best fit parameters for the HE and PA components
since they are not expected to be variable with time (see Figure~1).
The positronium continuum and narrow 511 keV annihilation line
integral fluxes
were $(2.0 \pm0.7) \times 10^{-3}$ and $(0.1 \pm0.3) \times 10^{-3} \,
{\rm photons \, cm^{-2} \, s^{-1} \, rad^{-1}}$, respectively.
The HE continuum 
intensity was determined from fits to these data combined with 
the collected Galactic plane observations following 
Kinzer et al. (1999). 
The HE component extrapolated to energies
below 1~MeV can be characterized as 
a power law function of photon index $1.6$ and normalization
of $3.6 \times 10^{-2} \, 
{\rm photons \, cm^{-2} \, s^{-1} \, MeV^{-1}\, rad^{-1}}$ at 100~keV.  

Modeling the OSSE data {\it alone} from 50-400~keV 
(obtained from the second viewing period), 
we find that the best fit is achieved
with an exponential cut off power law model 
\begin{equation}
A(E)=k(E/1\,{\rm keV})^{-\Gamma}\, \exp(-E/E_c)
\end{equation}
plus the HE continuum included as a fixed model component 
($\chi^2/\nu=3.95/9$). 
(We have subtracted the PA components from the OSSE data points according 
to the results obtained from the 4 week combined observations). 
The best fit parameters for the power law are: $\Gamma=-0.2^{+1.2}_{-2.5}$
and $E_c=30.1^{+47.3}_{-5.4}$ and a total flux (including the
HE continuum) of $7.8 \times 10^{-7} \, {\rm photons
 \, cm^{-2} \, s^{-1} \, \, keV^{-1} \, deg^{-2}}$ at 100~keV averaged
over the field of view of OSSE. 
All the quoted errors are for 90\% confidence limit. 
Removing the HE component produces a worse fit ($\chi^2/\nu=5.0/10$)
and yields a higher photon index and cut off energy that
are within the 90\% confidence limits of the previous fit.
Without the HE component, 
the $50-400$~keV data can also be satisfactorily fit ($\chi^2/\nu=9.4/10$)
with a single power law 
($A(E)=k(E/1\,{\rm keV})^{-\Gamma}$) of photon index $\Gamma=2.6\pm0.2$.

\subsubsection{RXTE/OSSE data: 3-400 keV}

We proceed by modeling the RXTE/OSSE spectra simultaneously using the
OSSE spectrum obtained in the {\it second} viewing period, but the PA component
subtracted from OSSE data. 
For the simultaneous fit, the flux normalization for the
two instruments must be handled consistently. 
For extended emission,
the measured flux depends on the field of view of each instrument
and the spatial distribution of
the diffuse emission. 
Hence, an {\it effective} solid angle (i.e. the
convolution of the distribution of the emission with the detector's response
function) should be calculated for each instrument.
In the case of {\sl OSSE} (fov of $11^\circ\! .4 \times 3^\circ\! .8$~FWHM),
convolving the soft $\gamma$-ray Galactic diffuse emission (assuming a
Gaussian latitude
distribution with $5^\circ$ FWHM)
and {\sl OSSE} 's triangular
response function yields an {\it effective} solid angle of
$\sim 1.3 \times 10^{-2}$~sr.
{\sl RXTE} 's solid angle is $\sim 3 \times 10^{-4}$~sr.
Hence, a composite data set was obtained by scaling down the OSSE data 
by a factor of $2.3 \times 10^{-2}$
to account for the difference in the solid angle of the two
instruments. 
In fitting the combined data, the intensity model parameters
for the two instruments has been set equal to each other.
In all the models presented hereafter, the diffuse HE continuum is 
included as a fixed model component present at all energies. 
It contributes approximately 13\% to the
total emission in the 3-100~keV band.
Generally, inclusion of this component tends to slightly
improve the fit. 

Examining the $10-400$~keV spectrum first, we find that the 
excess over the HE diffuse component is 
best described by an exponentially cut off power law model 
(equation~2) with the
following parameters ($\chi^2/\nu=24.8/43$):
photon index $\Gamma=0.63\pm0.25$, energy cutoff 
$E_c=41.4^{+13.0}_{-8.4}$ keV, and flux of 
$6.0 \times 10^{-7} \, {\rm photons \, cm^{-2} \, s^{-1} \, \, keV^{-1} 
\, deg^{-2}}$ 
at 100~keV. The total flux (including the HE continuum flux) at 100~keV 
is $7.7 \times 10^{-7} \, {\rm photons \, cm^{-2} 
\, s^{-1} \, keV^{-1} \, deg^{-2} }\, $. 
Notice that the best fit parameters are within the 90\% 
confidence limits of those derived
from fitting the OSSE data alone but that they are considerably 
more tightly constrained. 

As we discussed earlier, the flux intensity measured with OSSE depends on
the assumption of latitude distribution of the diffuse emission.
So far, we have presented results assuming a Gaussian distribution
of $5^\circ$ FWHM in latitude. To explore how the results depend on 
this assumption, we have   
simultaneously fitted the RXTE data with the OSSE data
scaled up and down by 50\%, respectively, to allow for
a different FWHM Gaussian distribution. Scaling up the OSSE
data by 50\% is equivalent to assuming a $ \sim 8^\circ$ FWHM 
Gaussian distribution while scaling it down  
by 50\% is equivalent to $\lesssim 0^\circ \! .1 $ FWHM.  
Scaling the OSSE data up yields a lower photon index 
($\Gamma=0.41^{+0.19}_{-0.17}$; $E_c=41.7^{+9.5}_{-6.3}$ keV) 
while scaling it down will have the opposite effect 
($\Gamma=1.23^{+0.33}_{-0.29}$; $E_c=51.6^{+37.8}_{-14.5}$ keV). 
In both cases,
the fit is worse and the fit parameters are marginally consistent
with the 90\% confidence limits of those parameters derived for a 
$5^\circ$ FWHM Gaussian assumption. 

Other models with high-energy cutoffs of different forms also provide good fits.
For the standard, $5^\circ$ FWHM Gaussian distribution, 
the standard pulsar model (equation~1)
yields $\Gamma=0.8^{+0.6}_{-0.4}$, $E_c=21.5^{+51.7}_{-21.5}$~keV,
and $E_f=45.3^{+20.3}_{-9.7}$ ($\chi^2/\nu=24.3/42$). 
The best-fit ($\chi^2/\nu=25.5/42$) 
Comptonized disk model (Sunyaev \& Titarchuk 1980)
has an input soft photon (Wien) temperature of
$< 1$~keV, a plasma electron temperature of 21~keV, and a plasma
optical depth of 3. 
A broken power law model
with photon indices of $1.4$ and $3.3$ and a break energy at
$75.2$~keV also provides a good fit ($\chi^2/\nu=29.0/42$),
but a simple power law model (best fit photon index of $1.7$)
does not ($\chi^2/\nu=97.3/44$).

\begin{figure}[h]
\centerline{
{\hfil
\psfig{figure=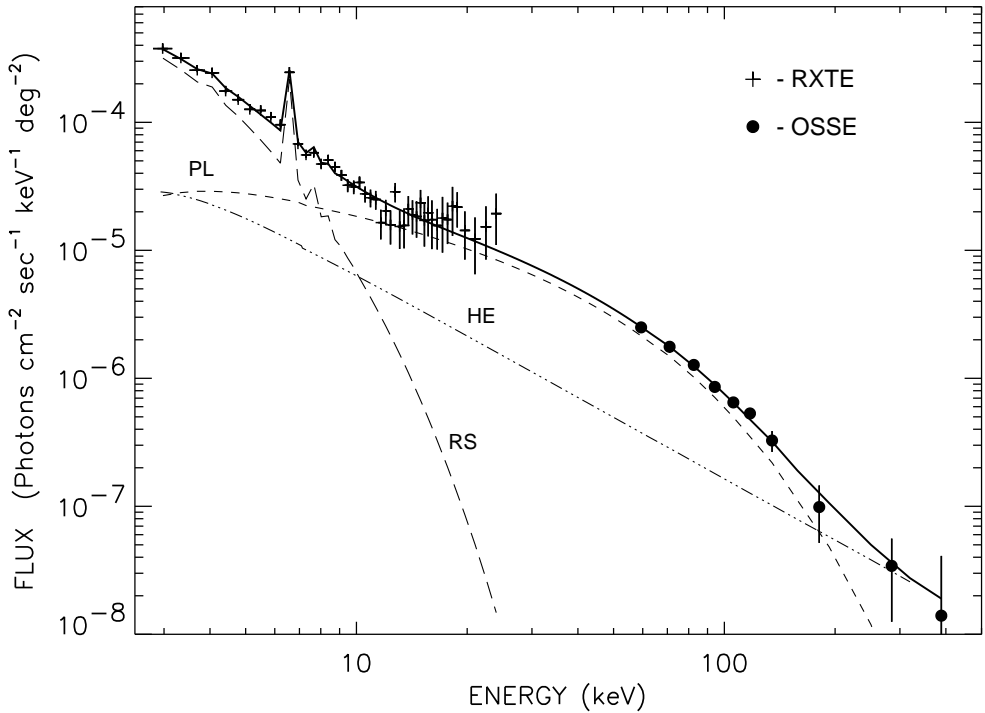,height=4.0truein,angle=90.0}
\hfil}
}

\caption
{\footnotesize Unfolded RXTE/OSSE spectrum and the best fit models in the 3-400 keV band.
OSSE spectrum
from weeks 3 and 4 of the observation is used. The positron annihilation
line and 3-photon positronium continuum components are subtracted from the
OSSE data. The combined spectrum (top
curve) is constructed  by 3 components: a thermal Raymond-Smith plasma model
(RS); an exponential cutoff
power law model (PL); and a high energy continuum (HE)
extrapolated to lower energies. Interstellar absorption is included as
part of the model.
See \S 3.2.2 for details.
} 
\end{figure}

We extended the RXTE/OSSE spectral fit down to 3~keV by 
adding a Raymond-Smith thermal plasma component to the 
exponentially cutoff power law model (and the HE continuum fixed in 
the model as before) and including the effect of Galactic absorption. 
The broader-range fit shown in Figure~2 yields 
$\chi^2/\nu=54.0/60$. The 3-10~keV spectrum is dominated 
(71\%) by the thermal plasma
component. The best model yields a hydrogen column density 
of $N_H=3.1\pm1.5 \times 10^{22} \, {\rm cm^{-2}}$ 
(the interstellar value for the line of sight at the center of the 
OSSE field of view    
is $\sim 2 \times 10^{22} \, {\rm cm^{-2}}$; the values for the lines of sight
to the FWHM corners of the field of view are as low as $0.4 \times 10^{22}
\, {\rm cm^{-2}}$), 
and a thermal plasma component of solar 
abundances and temperature $2.6^{+0.4}_{-0.3}$~keV with an unabsorbed
flux of $6.8 \times 10^{-6} \, {\rm photons \, cm^{-2} \, s^{-1} \, 
deg^{-2}}$ at
10~keV. The best fit parameters for the exponentially cutoff power law
model in this extended fit were $\Gamma=0.52\pm0.25$ and energy cutoff 
$E_c=39.4^{+11.2}_{-7.2}$ keV.

\section{SUMMARY AND DISCUSSION}  

We have measured the spectrum of the Galactic X-ray/$\gamma$-ray
background near the Scutum arm region. 
In addition to the extrapolated high
energy continuum (due to the interaction of cosmic rays with the 
interstellar medium) and the positron annihilation components, 
we have measured a variable hard X-ray/soft
$\gamma$-ray component dominating the 10-200~keV band.  
The shape of this component at its minimum observed intensity is best modeled by
an exponentially cut off power law of photon index of $\sim 0.6$ and
energy cut off of $\sim 41$~keV. 
We estimate that at 60 and 100~keV, known bright discrete sources
in the field of view of OSSE contribute about 46\% and 20\% to
the total (minimum) measured flux, respectively. 

The nature of the unresolved emission still remains to be determined. 
One interpretation is that the emission is due to a combination
of unknown discrete sources. 
In the course of PCA scans, we detected several low luminosity 
discrete sources of the order of a few mCrabs for which we do not 
have a positive identification with previously known sources. 
We were not able to  
determine the spectral shape of these sources above $\sim 20$~keV
because of the short exposures.
The integrated 2-20~keV flux from these weak sources is a substantial fraction of 
the total of the previously identified sources 
listed in Table 1 if GRS1915+105 is excluded.
It is then plausible that these weaker sources also make
a significant contribution to the flux detected with OSSE,
and in this case a significant fraction of the 40-100 keV flux observed with
OSSE would be due to discrete sources whose 2-10~keV flux
is brighter than $\sim 1$ mCrab.
The estimated 40-100~keV spectrum
of the bright sources listed in Table 1 is softer than the total OSSE 
spectrum measured during the second viewing period (see \S 3.1). 
The spectrum of the
remaining contributors would therefore have to be harder 
than that of the known sources reported in Table~1. 
While the spectrum during the second viewing period can be well
fit with the standard pulsar model, the best-fit parameters are unusual.
The HR of the model is higher than that of any pulsar in 
the review of White, Swank, \& Holt (1983), higher than ever seen
during the outburst of EXO~2030+375 (Reynolds, Parmar \& White 1993),
and higher than that of any of the pulsars in Table 1.
While the HR of EXO~2030+375 decreased as its luminosity decreased,
there is no clear dependence of the HR on luminosity for the pulsars
reviewed by White, Swank, \& Holt (1983).
If the observed spectrum is dominated by unresolved pulsars, they 
must have much harder spectra than typically seen for X-ray pulsars.
The spectra out to at least 40~keV is similar to that of the low 
state of black hole candidates (BHC) (e.g. Tanaka and Lewin 1995) since 
the unresolved spectrum can also be characterized with a broken power law
model of photon indices of 1.4 and 3.3 and a break energy at 75.2~keV.   
Higher quality data are needed to determine if the form of the
spectral break is similar to that of BHC. 

An alternative interpretation for the origin of the unresolved
emission is that it is truly of diffuse origin and is 
produced by mechanism(s) 
such as nonthermal electron bremsstrahlung or inverse Compton (IC) 
scattering of interstellar radiation off of cosmic-ray electrons
(e.g. Skibo \& Ramaty 1993). 
It is expected that the scale height of the emission due to IC
scattering be broad because of the large scale height of the
interstellar radiation field (optical, infrared, and the 
Cosmic microwave background). 
Indeed, the emission has been measured to be broad with
RXTE (Valinia \& Marshall 1998) and COMPTEL/CGRO (Strong et al. 1996b).
Therefore, in the diffuse  origin scenario,  
the hard spectral shape and the apparent broad extent
of the emission would suggest the possibility
that the IC scattering contribution may dominate over nonthermal
bremsstrahlung processes.  
Unlike the nonthermal bremsstrahlung scenario,
the IC scattering scenario has
the added advantage that the total power required is well within that
provided by Galactic supernovae. 

Our discovery of variability in the 40-100~keV flux from the Galactic
Ridge near Scutum shows that a substantial part of this emission is due to
discrete sources.
Determining the exact contribution of discrete sources and diffuse
mechanism(s) to the total Galactic plane emission requires instruments
capable of sensitive imaging over the hard X-ray/soft $\gamma$-ray band. 
A sensitivity of at least
$10^{-6}-10^{-7} \, {\rm photons \, cm^{-2} \, s^{-1} \, keV^{-1}}$ at
100~keV and spatial resolution of a few arc minutes are required. 

\acknowledgements
We thank David Band, Mark Finger, William Heindl, James Kurfess, 
Ed Morgan, and Mike Muno
for helpful discussions. We also 
thank Keith Jahoda for help with the scanning response matrix for RXTE.

%

\newpage
\small{
\begin{deluxetable}{lcccrrrr}
\tablecolumns{8}
\tablecaption{Bright Discrete Sources Within OSSE Field of View}
\tablewidth{0pc}
\tablehead{ Name  & \multicolumn{2}{c}{Flux (2-10 keV)\tablenotemark{a}} &
OSSE\tablenotemark{b}  & \multicolumn{2}{c}{Flux at 60 keV\tablenotemark{c}}  &
\multicolumn{2}{c}{Flux at 100 keV\tablenotemark{d}}   \\
  &  \multicolumn{2}{c}{mCrab}  &   Response(\%) &
 \multicolumn{2}{c}{$10^{-5} {\rm photons \, cm^{-2} \, s^{-1}}$} &
\multicolumn{2}{c}{$ 10^{-6} {\rm photons \, cm^{-2} \, s^{-1}}$} \\
   &   W1-2\tablenotemark{e} &  W3-4\tablenotemark{f} & &
W1-2\tablenotemark{e,g}  & W3-4\tablenotemark{f,h} & W1-2\tablenotemark{e,i} &
W3-4\tablenotemark{f,j} }
\startdata
A1845-024  & $9.1$ & $3.9$  & 69  &  0.7 (4\%) &  0.3 (3\%) & 1.5 (4\%) & 0.6 (2\%)
   \nl
XTE J1858+034 &  $18.5$ & $15.9$  &  66  &  1.2 (6\%) & 1.1 (10\%) &
1.2 (3\%) & 1.1 (3\%)   \nl
GS 1843+009  & $12.3$ &  $6.4$  &  54  &  5.1 (27\%) & 2.7 (24\%)   & 6.7 (16\%) &
3.5 (11\%)   \nl
XTE J1855-026  & $8.1$ &  $5.3$  &   36  & 1.6 (9\%) &  1.0 (9\%) & 1.9 (5\%) &
1.2 (4\%)   \nl
SS433  &   $10.3$  & $5.5$  &  17  & 0.1 ($<1$\%) &  0.06 ($<1$\%) & 0.07 ($<1$\%)
& 0.04 ($<1$\%) \nl
GRS1915+105  &   $620$ &  $521$   &  $<1$  & $\sim 0 (0\%)$   & $\sim 0 (0\%)$   &
$\sim 0 (0\%)$  & $\sim 0 (0\%)$ \nl \cline{1-8}
Total     &    &    &    &   8.7 (46\%)  &  5.2 (46\%)
&   11.4 (28\%)  &   6.4 (20\%)  \nl \cline{1-8} 
Residual Flux\tablenotemark{k} &  &  &  &  10.1 &  5.8 &  29.4 &  26.3 \nl \cline{1-8}
\enddata
\tablenotetext{a} { Source flux from RXTE All Sky Monitor (ASM). Estimated
RMS errors are typically 5\%-20\%.}
\tablenotetext{b} { OSSE's response at the position of the source.}
\tablenotetext{c} { Estimated source flux through OSSE's collimator
(and the $\sim$\% contribution of the sources to the total OSSE flux
in each observation at 60~keV).
This number has been determined by multiplying OSSE's response at the
position of the source by the estimated flux of the source as described
in the text.}
\tablenotetext{d} { Same as (c) except at 100~keV. }
\tablenotetext{e} { Averaged over weeks 1 and 2 of the observation.}
\tablenotetext{f} { Averaged over weeks 3 and 4 of the observation.}
\tablenotetext{g} { Total flux through OSSE's collimator averaged over
                    weeks 1 and 2 was $1.88 \times 10^{-4} \, {\rm photons 
                    \, cm^{-2} \, s^{-1}}$ at 60~keV.}
\tablenotetext{h} { Total flux through OSSE's collimator averaged over 
                    weeks 3 and 4 was $1.10 \times 10^{-4} \, {\rm photons 
                    \, cm^{-2} \, s^{-1}}$ at 60~keV.}
\tablenotetext{i} { Total flux through OSSE's collimator averaged over
                    weeks 1 and 2 was $4.08 \times 10^{-5} \, {\rm photons 
                    \, cm^{-2} \, s^{-1}}$ at 100~keV.}
\tablenotetext{j} { Total flux through OSSE's collimator averaged over 
                    weeks 3 and 4 was $3.27 \times 10^{-5} \, {\rm photons 
                    \, cm^{-2} \, s^{-1}}$ at 100~keV.}
\tablenotetext{k} {Total measured OSSE flux minus the estimated
                   contribution of discrete sources.} 

\end{deluxetable}

}

\end{document}